# The Discovery of European Porcelain Technology

C.M. Queiroz[1], S. Agathopoulos

Ceramics and Glass Engineering Department, University of Aveiro,
University Campus, 3810-193 Aveiro, Portugal
[1]corresponding author; email: queiros@cv.ua.pt

**Abstract**

The European quest for the hard paste porcelain lasted until about 1710, when the first production unit was built at Meissen in Saxony. Although, it is generally believed that Böttger discovered the European porcelain technology, this work shows that the German porcelain quest was a State affair; a collective project directed by Tschirnhaus and supported by August the Strong. It also aims to shed light to the reasons that led to Tschirnhaus's relative oblivion. Considering three distinct temporal stages, we re--examine the roles of the main contributors to the porcelain project, referring to the role of the Mercantilism system and to the philosophical roots of Tschirnhaus's method.

**1. The Trigger Conditions: Mercantilism, Baroque and Modern Science**

The German discovery of porcelain production techniques was probably the most emblematic technical achievement of Baroque period, aimed to fulfil the Baroque aesthetical opulence and the luxurious way of life of the noble class, under the economical constraints of the Mercantilist policy. This quest should be considered as one of the earliest successes of modern experimental method, as outlined by Francis Bacon in *Novum Organum* (1620), emphasizing the experimentalist approach, while the importance of mathematics was underestimated. Bacon's philosophy was highly

*1*

influential on scientists like Robert Boyle and Robert Hooke (Solís, 1985). This approach opposes to Cartesianism, defending that the route for scientific knowledge should rely on mathematical reasoning. Along with Huygens and Leibniz, Ehrenfried Walter von Tschirnhaus (1651-1708) tried to conciliate these two apparently opposing methods. This development led to modern science (Hooykaas, 1986) and was crucial to the success of the porcelain research project.

**2. Chronological Stages of Hard-Paste Porcelain Research Project**

August the Strong (1670-1733), Elector of Saxony and King of Poland, gave Tschirnhaus the assignment of searching the whole Saxon territory for deposits of precious stones, while developing methods of polishing and lapping these valuable resources. At the same time, he pressed Johann Friedrich Böttger (1682-1719) to make gold. These two projects merged in the hard-porcelain industrial project, were three conceptually distinct stages can be distinguish.

2.1. *Laboratory research: technology, raw materials and eutectics (1682-1706)*

Tschirnhaus was born in April 1651 in Kieslingswalde (Germany). He studied in Leiden University from 1668 to 1674. In 1673-74, he was introduced to Spinoza in Leiden. In 1675 he began a "knight's tour", meeting Papin, Boyle, Oldenburg, John Wallis and John Collins in Britain. In the same year, he met Leibniz and Huygens, and continued his learning travel in Paris, where he met Colbert[1]. In 1676, he went to southern France and Italy, returning to his estate in Kieslingswalde in 1679 (Leinsle, 1997). He went again to France in 1701-02, where he visited the St. Cloud soft-paste porcelain factory (Raffo, 1982), becoming a member of the Acadèmie Royale des Sciences (Paris) in 1682. He developed the idea of the worthiness of matching mathematics and physics in his chief philosophical work, *Medicina Mentis* (1687).

Through the 1680's, he applied his scientific knowledge to the development of industrial processes, focusing at the manufacture of glass and porcelain. He also established the first Saxon glass manufactures and developed polishing technology (Bauke, 1990). By 1682, he studied theoretically the envelope of light beams emitted from a point source after reflection on a parabolic surface (O'Connor and Robertson, 1997), as a preliminary step towards the development of large burning lens and mirrors (Hofmann, 1981). The last ones were an improved version of the mirrors developed earlier by François Villette, in France (Klinckowstroen, 1965). The use of such equipments allowed him to reach temperatures within 1500-2000ºC, higher than what he could achieve in contemporary combustion furnaces (Bauke, 1990). The method was welcome in laboratory research because the samples could be easily observed and many trials could run in a short time. From 1693-94 onwards, Tschirnhaus found that porcelain could melt, and studied the refractory properties of several possible raw minerals (Leinsle, 1997). Understanding eutectics, he discovered that porcelain could be obtained from a mixture of clay and fusible minerals (flux). Tschirnhaus conducted most of this research guided by his prior knowledge of glass technology, whereby he seemingly believed that porcelain was a glass$^2$. By 1699, he actually obtained "porcelain" by melting and cooling mixtures of raw materials (Bauke, 1990).

Tschirnhaus started the porcelain research long before Böttger joined the quest, as his assistant, in 1705 (Goder and Walter, 1982). Böttger was born in 1682 at Schleiz. After he started working as a pharmacist apprentice in Berlin, he devoted himself to alchemy (Hofmann, 1982). In 1701, he made his first demonstration in the presence of trustworthy witnesses. He pretended to transmute two silver coins into gold. After his "success", the Prussian King, tried to lock him as a treasure, but he managed to escape to Saxony. However, his fame preceded him, and August the Strong kept him watched and confined in Albrechtsburg Castle (Meissen), trying to exploit his "gold maker" skills. Later on, as no gold was forthcoming, August directed him to porcelain research.

2.2. *Scale-up: from laboratory to manufacture (1706-1710)*

After the preliminary laboratory trials provided promising porcelain samples, a scale-up procedure began. It started in 1706, when August ordered metallurgist experts to be hired in the Freiberg region, since they held valuable knowledge on thermal processes and kilns. The Freiberg Mining Counsellor, Gottfried Pabst von Ohain (1656-1729), expert in mineralogy and metallurgy, was chosen to direct the iron-metallurgist workers hired in January 1706 (Goder and Walter, 1982, p.75). Tschirnhaus was godfather of Ohain's son (Westfall, 1995); therefore, he supported Ohain's cooperation. Ohain, contributed with his vast practical knowledge of ores and minerals, as an expert furnace constructor, and developed formulas for overglaze enamels. Similarly to Tschirnhaus, his contribution has been largely disregarded. Ohain provided Böttger with selected mineral samples, which were otherwise inaccessible to him.

In September 1707, three cellar vaults were dedicated to porcelain production at the Dresden fortress, providing an intermediate level between laboratory and factory's production scales. The production of red stoneware was soon started, from iron-rich refractory clay mixed with a fusible flux fired to very high temperature. This stoneware was however extremely hard. Thus, it was cut, polished and wheel-engraved like a semi-precious stone, with the aid of equipment developed by Tschirnhaus. This period was characterized by the decline of Tschirnhaus's influence, and by the abandon of the former Tchirnhaus's program, based on the use of sun light kilns, which were appropriate for laboratory trials but unsuitable to be scaled-up to the factory's level. This decay was compensated by the rising influence of the two new members who joined the porcelain "contubernium" (task force), Pabst von Ohain and Dr. Bartholomaei (1670-1742). The latter was a medicine physician and naturalist who, working from 1707 onwards carried out numerous trials of paste preparation, intended to find the most favourable clay raw materials among those found in Saxony region (Raffo, 1982). The relative shares of the chosen clays were mostly established by him. A protocol dated from 15.01.1708 certifies the invention of the European hard paste porcelain, establishing the transition from the laboratory trial stage to the production scale maturity. This document has been attributed either to Böttger (Hofmann, 1980, p. 35) or to Bartholomaei (Bauke, 1990).

In 1708-09, the first Saxon white clay (kaolin) was found in Erzgebirge. This discovery is often attributed to Bötger, but the research was actually conducted by Ohain (Bauke, 1990) and Bartholomaei (Goder and Walter, 1982). This finding was crucial for the success of the manufacture of white porcelain. New deposits were found later at Aue, near Schneeberg, in Saxony, and a mixture of the two clays with calcified alabaster and white silica was found particularly successful (Raffo, 1982, p. 84). After Tschirnhaus' death (14.10.1708), Böttger wrote a report to the Elector (28.03.1709), claiming that he had prepared "good white porcelain" equivalent or even better than the Chinese one (Raffo, 1982). The official start-up date of the porcelain manufacture (at Dresden) dates from 23.01.1710, when the Saxon Court Chancellery issued a "Supreme Decree" in Latin, French, German and Dutch, announcing the discovery to the World. For security reasons, a Decree from 7.03.1710 transferred the factory to Albrechtsburg (Meissen), being officially opened in August 6, the same year (Hofmann, 1980, p. 36).

2.3. *Full scale production and improvement of decorating techniques (1710 onwards)*

Although the porcelain manufactory at Meissen started the production of porcelain in 1710, the decoration was poor, and the first significant sales only took place at the Leipzig Easter Fair in 1713 (Goder and Walter, 1982). The nine years held between the manufacture start-up and Böttger's death, in March 1719, were characterized by a constant search for improvement of the manufacturing techniques. A wide variety of enamel pigments were developed, obtained from powdered metallic oxides.

**3. The Priority Dispute: Böttger or Tschirnhaus?**

The inspector of Meissen manufactory, Johann Melchior Steinbrück (1673-1723), was largely responsible for the initial promotion of Böttger as a genius, to whom he claimed that all credit of the discovery should be given (Bárta, 1963). He stated, in a laudatory text (Steinbrück, 1713), that his «*inventoribus rerum*» would hardly be repeated in a hundred years! Steinbrück was a close relative of Böttger (his brother in law), and under Steinbrück's suggestion (1715), Augustus chose Böttger as director of porcelain manufacturing and attributed him the title of Baron. The secrecy policy followed by Augustus also contributed for the relative oblivion of Tschirnhaus's role, since early documents were secret. Earlier historians, following Steinbrück, emphasized Böttger's contribution instead of Tschirnhaus's, starting a chain that still echoes.

Bárta (1963) pointed that religious intolerance could have also played a role on Tschirnhaus's oblivion. Some of his ancestors were considered to be sympathetic with the catholic religion, particularly his father's uncle living in Prague (visited by his father in 1637), since the catholic religion was sustained by the majority of the Bohemians (while in Saxony the Lutheranism was officially supported). It can be argued, however, that this Catholic's influence suspicion did not hinder his earlier fair relationship with the Court. Therefore, a more important contribution should be sought in the reaction of the orthodox Lutheranism against the Enlightenment. This contribution probably was reinforced by a similar reaction from the influential secret brotherhood of the Rosy Cross[3].

Tschirnhaus was considered as one of the most influential leaders of the German Enlightenment (Winter, 1960), a cultural movement that criticized the more orthodox Lutheranism (Wolter, 1971). His mind echoed the new tendencies of European philosophy, as Cartesianism (Künzel, 1989a, b) and Spinozism (Künzel, 1989b; Wurtz, 1991; Vermij, 1991). These tendencies were not welcome in the orthodox Lutheranism circles (Wolter, 1971). Particularly, Wolff (a Leibniz' disciple) was dismissed from Halle University (1721), and later, Christian Thomasius accused Tschirnhaus of Spinozism (*Monatsgesprächen*, 1688), and therewith of atheism (Leinsle, 1997; Wurtz, 1980). As a symptomatic consequence of this opposition, we see that from 1704 onwards, Tschirnhaus held a dispute with the preacher Kellner von Zinnendorf (1665-1738) (Sabelleck, 1998; Knobloch, 2003).

As regards to Böttger, his alchemistic skills have been seemingly better tolerated by the Lutheran orthodoxy, since alchemy was relatively well accepted in Protestant circles.

**Notes**

[1] Jean-Baptiste Colbert (1619-1683), minister of Louis XIV, leader of the Mercantilism.

[2] We know now that porcelain is only partially vitrified. The glass phase is enforced with long crystals of mullite, knitting the whole porcelain structure together.

[3] The brotherhood of the Rosy Cross; a secret society blending alchemy with protestant religion, which attracted many adepts in Germany (possibly including Böttger himself ).